\documentclass[12pt]{article}

\usepackage{scicite}
\usepackage{times}
\usepackage[dvipsnames]{xcolor}
\usepackage{graphicx}
\usepackage{bbold}
\usepackage[normalem]{ulem}

\newcommand{\eq}[1]{Eq.~(\ref{#1})}

\newcommand{\fig}[1]{Fig.\,\ref{#1}}

\def\be{\begin{equation}}
	\def\ee{\end{equation}}
\def\bea{\begin{eqnarray}}
	\def\eea{\end{eqnarray}}

\makeatletter
\newcommand*\bigcdot{\mathpalette\bigcdot@{.5}}
\newcommand*\bigcdot@[2]{\mathbin{\vcenter{\hbox{\scalebox{#2}{$\m@th#1\bullet$}}}}}
\makeatother

\newenvironment{sciabstract}{%
\begin{quote} \bf}
{\end{quote}}

\title{Strong Interlayer Coupling and Stable Topological Flat Bands in Twisted Bilayer Photonic Moir\'e Superlattices}

\author
{Chang-Hwan Yi,$^{\ast}$ Hee Chul Park,$^{\dagger a}$ Moon Jip Park$^{\dagger b}$\\
\\
\normalsize{Center for Theoretical Physics of Complex Systems, Institute for Basic Science (IBS), }\\
\normalsize{Daejeon, 34126, Republic of Korea}\\

\normalsize{E-mail: $^\ast$yichanghwan@hanmail.net, $^{\dagger a}$hcpark@ibs.re.kr, $^{\dagger b}$moonjippark@ibs.re.kr}\\
\normalsize{$^\dagger$Corresponding authors equally contributing to this work.}
}
\date{}

\begin{document} 


\baselineskip24pt


\maketitle


\begin{sciabstract}
    The moir\'{e} superlattice of misaligned atomic bilayers paves the way for designing a new class of materials with wide tunability. In this work, we propose a photonic analog of the moir\'{e} superlattice based on dielectric resonator quasi-atoms. In sharp contrast to van der Waals materials with weak interlayer coupling, we realize the strong coupling regime in a moir\'{e} superlattice, characterized by cascades of robust flat bands at large twist-angles. Surprisingly, we find that these flat bands are characterized by a non-trivial band topology, the origin of which is the moir\'{e} pattern of the resonator arrangement. The physical manifestation of the flat band topology is a robust one-dimensional conducting channel on the edge, protected by the reflection symmetry of the moir\'{e} superlattice. By explicitly breaking the underlying reflection symmetry on the boundary terminations, we show that the first-order topological edge modes naturally deform into higher-order topological corner modes. Our work pioneers the physics of topological phases in the designable platform of photonic moiré superlattices beyond the weakly coupled regime.
\end{sciabstract}

\section*{INTRODUCTION}
    When two sheets of atomic bilayers are stacked with a finite rotation angle, the periodicity of the two incommensurate layers produces a large moir\'{e} superlattice. This giant amplification of the crystalline periodicity is the hallmark of moir\'{e} materials~\cite{Andrei2020} and provides a viable platform for  band structure engineering. Magic-angle twisted bilayer graphene is the representative example, exhibiting a variety of novel quantum phases such as superconductivity~\cite{Cao2018}, correlated insulators~\cite{Cao2018i,Lu2019,Kerelsky2019}, and topological phases~\cite{PhysRevLett.123.036401,PhysRevLett.123.216803,PhysRevLett.126.066401}. At the microscopic level, van der Waals coupling between the two layers is a weak interaction, but it is the key ingredient that drives the drastic deformation of the band structure~\cite{Bistritzer12233}. 
    
    A photonic resonator array provides an attractive platform to explore the physics of moir\'{e} materials. Realizing a moir\'{e} superlattice in photonic crystals has clear advantages over electronic systems as the interactions in photonic crystals are not limited to weak van der Waals coupling~\cite{RevModPhys.91.015006,chen2021perspective,PhysRevLett.126.136101,PhysRevLett.126.223601}. The tunable geometry and dielectricity of the photonic crystals allow a feasible control of the interlayer couplings. Indeed, photonic systems can expand the scope of moir\'{e} materials beyond the weak coupling regime. For instance, by employing the interlayer distance parameter, the Dirac point-associated flat bands and the corner states in photonic systems~\cite{oudich2021photonic}, as well as the magic angle analogies in the acoustic moir\'e systems~\cite{deng2020magic}, have been examined.
	
    Although the photonic moiré superlattice directly analogies with the twisted bilayer graphene~\cite{Cao2018i,Lu2019,Kerelsky2019}, we find a drastically different band structure distinct from the standard phenomenology of its electronic counterpart. For the first time, we discover the robust topological flat bands at arbitrary large twist-angles, resulting from particular localization patterns associated with the Aharonov-Bohm cage effect. This localization is unique to the strong coupling regime of the moiré superlattice. We emphasize that our discovery of the intrinsic relationship between the topology and flat bands differentiates our works from the previous studies in~\cite{Wang2020,fu2020optical}. Physical manifestations of our flat bands include one-dimensional helical edge modes and higher-order topological corner modes. Our results can provide a novel platform to design photonic topological materials and topological waveguides~\cite{photonic_wiersig,photonic_EP} utilizing the moiré superlattice.

\begin{figure}[t!]
\centering
	\includegraphics[width=0.5\textwidth]{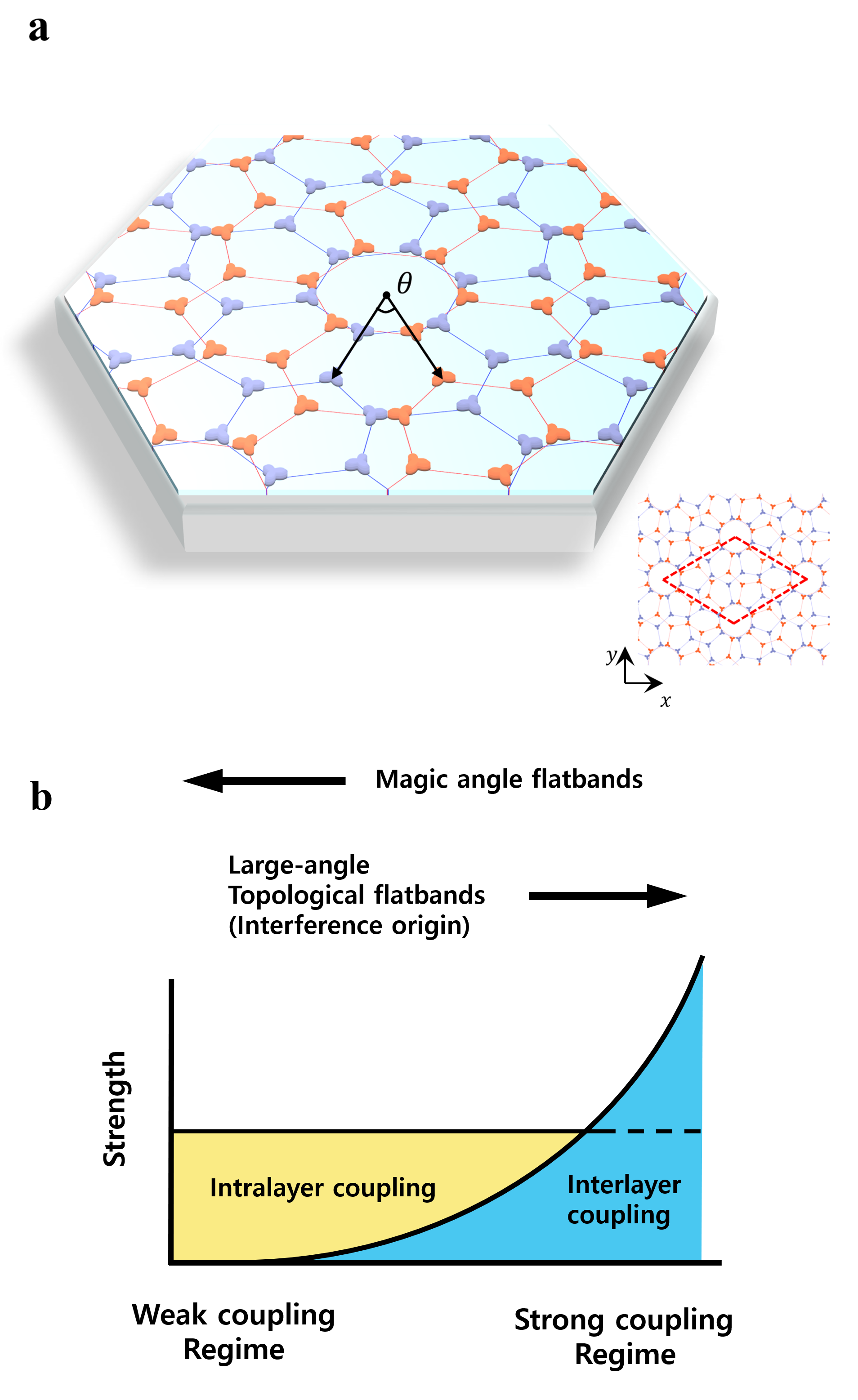}% Here is how to import EPS art
	\caption{\textbf{a} Picture of the photonic moir\'{e} superlattice. Two honeycomb lattices of dielectric resonator quasi-atoms (red and blue) are overlaid with twist-angle $\theta$. The primitive unit cell in the right panel is marked with the red rhombus and contains 28 quasi-atoms.
		\textbf{b} Schematic illustration of the strong coupling regime of the moir\'{e} superlattice. As the interlayer couplings exceed the intralayer couplings, we find topological flat bands at large twist-angles. This feature is in direct contrast with the weak coupling regime, where flat bands appear only at the magic angles. See the sections S14-18 in Supplementary Materials for the tunability of the interlayer couplings as a function of the refractive index and the twist-angle.}
	\label{fig:1}
\end{figure}	

\section*{RESULTS}
\paragraph*{Photonic moir\'{e} superlattice\\}
    Our photonic moir\'{e} superlattice is constructed by twisting two honeycomb photonic crystals with AA stacking around the hexagonal center. Figure~\ref{fig:1}\textbf{a} exemplifies our system setup with a rotation angle of $\theta=21.78^\circ$, which forms the smallest possible moir\'{e} superlattice~\cite{PhysRevLett.99.256802,PhysRevB.98.085435}. Each photonic crystal site consists of a quasi-atom established upon a dielectric resonator having a radius $a/6$, where $a\equiv1$ denotes a resonator-resonator lattice constant about the resonator center positions (i.e., the face-to-face resonator-wise gap is $2a/3$). If not mentioned otherwise, we fix a refractive index inside the resonator as $n=4$ and outside as $n=1$. A set of $28$ quasi-atoms forms a single moir\'{e} unit cell, as shown in the right-lower small panel of Fig.~\ref{fig:1}\textbf{a}. To obtain the energy bands of the optical modes, we solve the two-dimensional Helmholtz wave equation,
\bea
	-\nabla^2\psi=n^2(\mathbf{r})\frac{\omega^2}{c^2}\psi\ ,
	\label{eq:helm}
\eea
    imposing proper periodic boundary conditions. Throughout the whole of this work, we numerically tackle \eq{eq:helm} by employing the so-called boundary element method~\cite{BEM_wiersig,BEM_knipp}. The details on the implementation of this method are given in the METHOD section and Supplementary Materials. Here, $n(\mathbf{r})$ is the piecewise constant refractive index, and $\omega =c k$ is the free-space temporal frequency with a vacuum wavenumber $k$. In a single circular-shaped dielectric resonator, the resonator mode comprises multiple quasi-atomic orbitals with different azimuthal orbital numbers, $l$, such that $\Psi(r,\theta)\sim \psi(r)\exp{(il\theta)}$~\cite{vahala2,optical_process}.
    \begin{figure*}[t!]
\centering
	\includegraphics[width=1\textwidth]{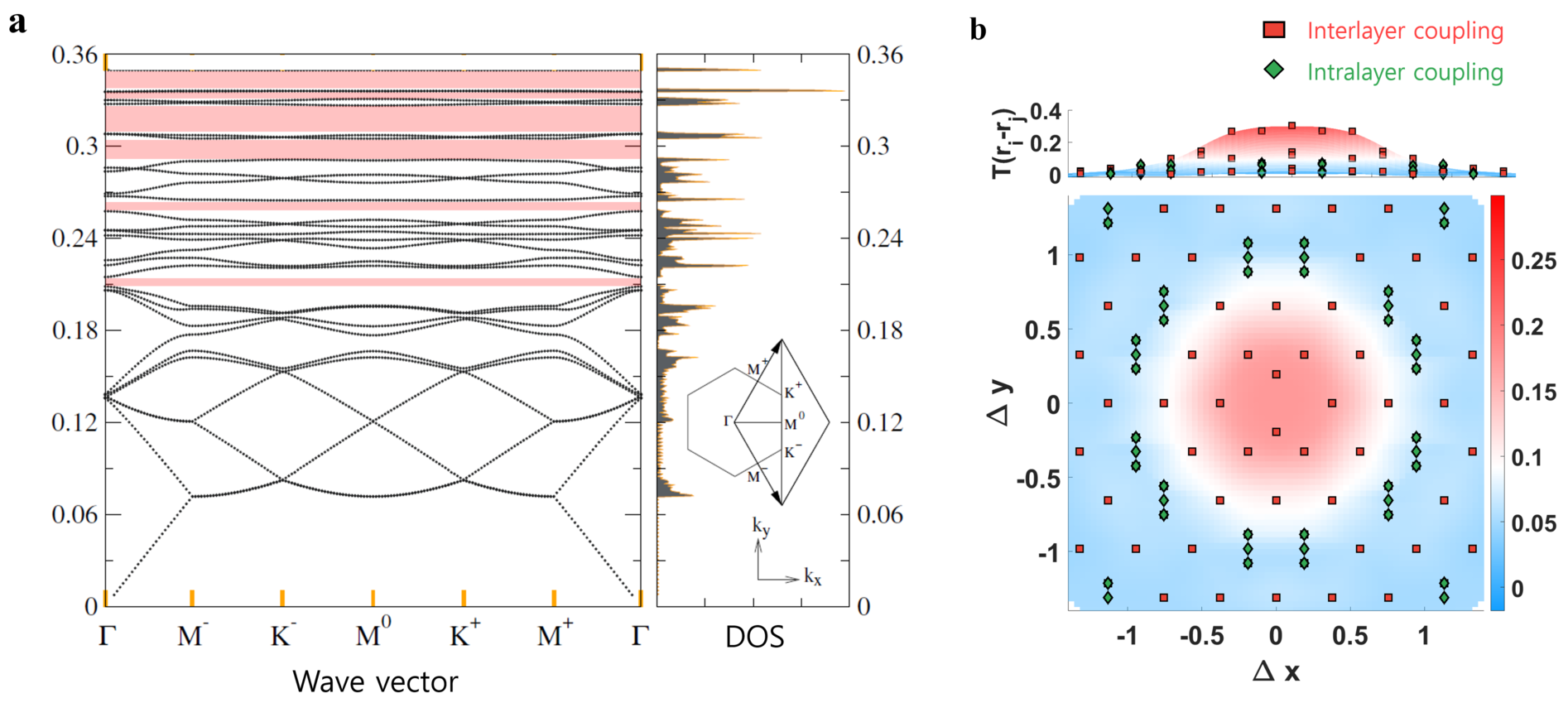}% Here is how to import EPS art
	\caption{
		\textbf{a} Band structure of the lowest 28 photonic energy bands. Contrary to the weak coupling regime of the moir\'{e} superlattice, we find large gap openings in the high energy modes (pink regions).
		\textbf{b} Hopping strength as a function of the displacement between two different quasi-atoms along the $x$ and $y$ directions. We find that the strength of the interlayer coupling (red squares) dominates over the intralayer coupling (green diamonds). This result directly verifies that our photonic crystal exhibits the strong coupling regime of the moir\'{e} superlattice. }
	\label{fig:11}
\end{figure*}	

    Given the moir\'{e} superlattice of optical resonators, we find that a number of energy bands stemming from different azimuthal orbital number sectors severely overlap one another in the low $n$ regime (e.g., $n\sim 2$, etc.). This is due to the weak confinement inside the resonators that enhances the resonator--resonator mode couplings. Contrarily, as $n$ increases (up to 4 in our case), these couplings are suppressed to the tunneling regime, where the whole band structure is energetically separated into a set of 28 distinct bands (equivalent to the number of quasi-atoms in a single unit cell). In \fig{fig:11}\textbf{a}, we show those of the lowest 28 bands corresponding to linear combinations of the single quasi-atomic modes with $l=0$. The following sequential bundles of 28 bands are from combinations of the single atomic modes with $l\ge 1$. The detailed wave characters of each band are shown in Figs. S6, S7, and S8 in Supplementary Materials. In addition, the mixed band structures in the lower refractive index regime and the separated band structures in the higher refractive index regime are shown in Fig. S11 in Supplementary Materials.

    As the bands with $l=0$ and $l\ge1$ are energetically decoupled in this high refractive index regime, the localized Wannier orbitals can be reconstructed from the solution of the Helmholtz equation. The effective tight-binding model that exactly reproduces the energy dispersion of the photonic bands can be obtained as 
\bea
H =\sum_{i,j}
T(\mathbf{r}_i - \mathbf{r}_j)
|\mathbf{r}_i\rangle\langle \mathbf{r}_j|+ {\rm h.c.}\ ,
\eea
    where $|\mathbf{r}_i\rangle$ is the $i$-th localized quasi-atomic state at site $\mathbf{r}_i$, and $T(\mathbf{r}_i - \mathbf{r}_j)$ is the effective hopping strength between sites $i$ and $j$ (see Supplementary Materials for the detailed algorithm of the Wannierization). Figure \ref{fig:11}\textbf{b} shows $T(\mathbf{r}_i - \mathbf{r}_j)$ as a function of the spatial displacement between two quasi-atoms, where the red squares and green diamonds represent the coupling between interlayer and intralayer sites, respectively. We find that the strength of the interlayer coupling dominates over that of the intralayer coupling. As a result, our analysis directly verifies the strongly coupled regime of the moir\'{e} superlattice.

\begin{figure*}[t!]
\centering
	\includegraphics[width=0.95\textwidth]{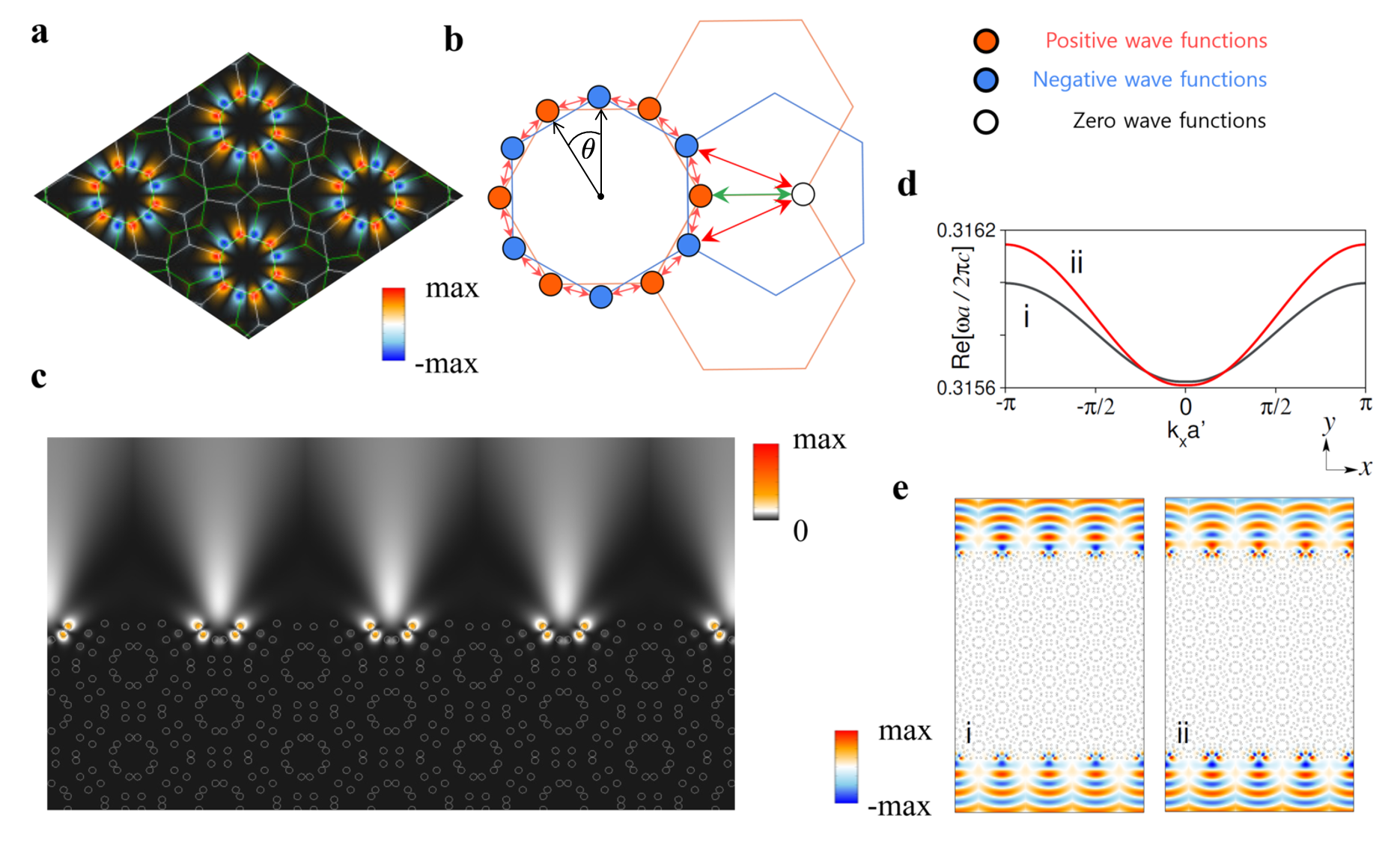}% Here is how to import EPS art
	\caption{\textbf{a} Wave function of the flat bands. We find that the wave functions form localized standing waves around the innermost dodecagonal quasi-atoms near the rotational center with sign oscillations. \textbf{b} Illustration of the destructive interference effect in the wave functions of the flat bands. The alternating signs of the standing waves introduce destructive interference in the hopping to outer quasi-atoms. \textbf{c} Intensity ($|\psi|^2$) pattern of the edge modes magnified along the upper edge of the lattice. \textbf{d} and \textbf{e} Energy dispersion of two edge modes and corresponding wave functions of Re$(\psi)$ obtained for $k_x=0$. Wave functions i and ii represent the symmetric and antisymmetric modes, respectively.}
	\label{fig:2}
\end{figure*}

\paragraph*{Flat bands in the strongly coupled regime\\}
    We focus on the set of the bands with $l=0$. First of all, in the low-energy bands, robust gap closings are observed at the $\textrm{K}^+$ and $\textrm{K}^-$ points in the Brillouin zone (BZ). Such gap closings are reminiscent of the Dirac cones that have been previously observed in monolayer honeycomb lattices and are protected by $\pi$ Berry phases. We additionally discover large gap openings (pink bands in Fig. \ref{fig:11}\textbf{a}) separating multiple flat bands at higher energies ($\omega a/2\pi c > 0.2$). The large gap openings are a distinct feature compared to the gapless spectrum of a single-layer honeycomb lattice. Furthermore, we find that the wave characters of the emergent flat bands form anomalously localized states within the innermost dodecagonal quasi-atoms near the rotation center (see Fig. \ref{fig:2}\textbf{a} and Supplementary Materials for further details).

    The localized wave functions of the flat bands can be represented as standing waves with sign oscillations:
\bea
\Psi \sim \sum_{j} (-1)^{\alpha_j} \psi_{j}\ ,
\eea
    where $\psi_j$ is the localized wave function in the innermost dodecagonal quasi-atoms with $j\in \{ 1,2,3,...,12 \}$, and $\alpha_j$ is an integer describing different angular wavenumbers of the standing waves. The couplings marked by red colors in Fig. \ref{fig:11}\textbf{b} represent the dominant couplings between the innermost dodecagonal quasi-atoms. Although we still find finite couplings to other sites, the standing waves cannot propagate further to the outer quasi-atomic sites, as the sign change in the wave functions produces destructive interference to the outer sites (red and green arrows in Fig. \ref{fig:2}\textbf{b}). As a result, the wave functions of the innermost quasi-atoms form a localized standing wave (LSW) with flat bands . In our works, it turns out that the destructive interference is universally observed at arbitrary large twist-angles. This particular type of localization is only possible if the strengths of the interlayer and intralayer hoppings are comparable, and it serves as the hallmark of the strong coupling regime of the moir\'{e} superlattice. An explicit demonstration of the competing characteristics between the intralayer and interlayer coupling strength as a function of the twist-angle is given in Supplementary Materials: It is found that the Ahronov-Bohm destructive interference is most prominently observed in the twist angle $\sim$13.17\textdegree~ where the interlayer coupling cancels each other (see Fig. S13 for band structures for different twist angles and see S15 for the coupling strength competition).

\paragraph*{Non-trivial topology induced by a moir\'{e} pattern\\}

\begin{figure*}[t!]
\centering
	\includegraphics[width=1\textwidth]{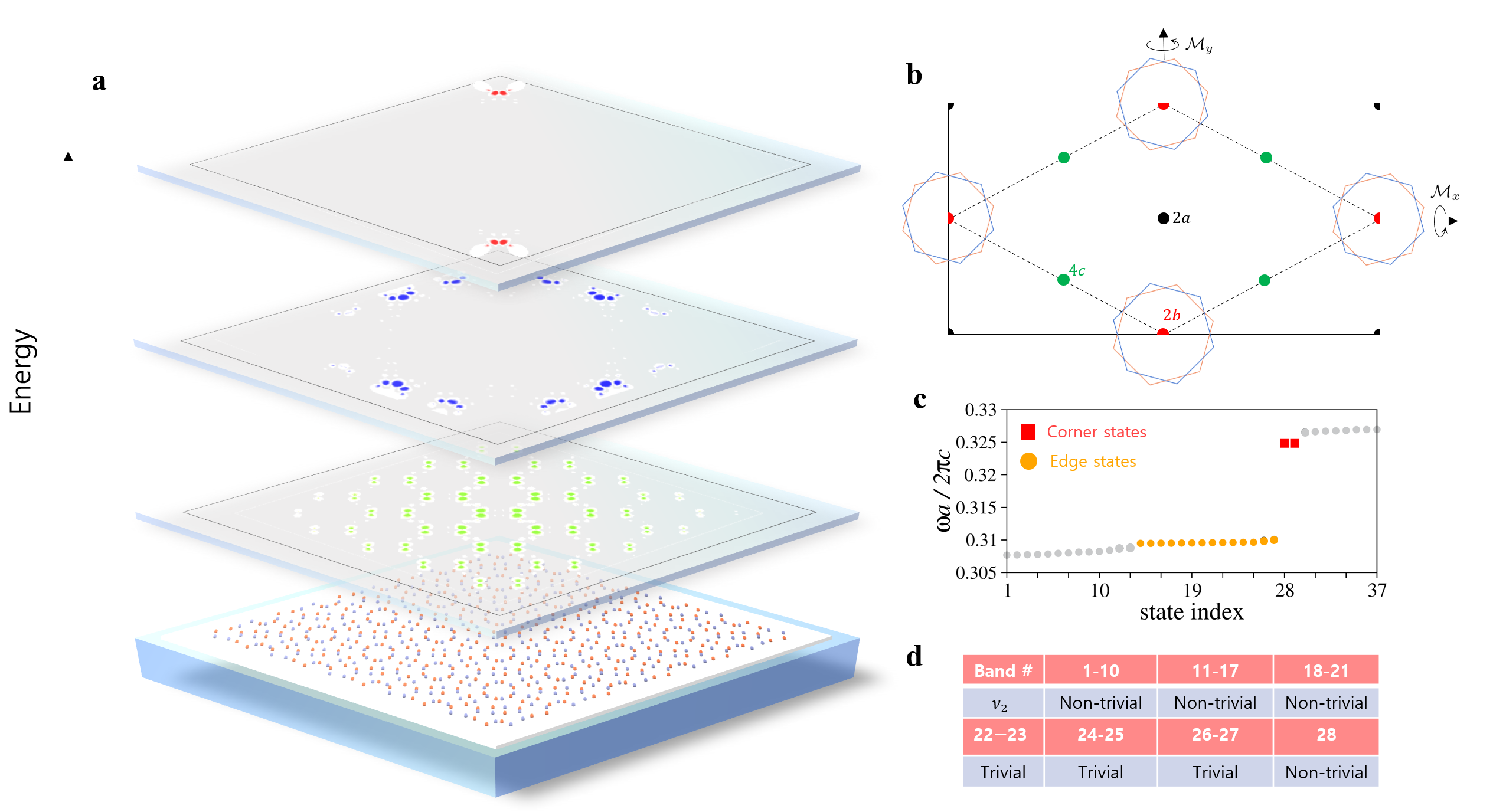}% Here is how to import EPS art
	\caption{ \textbf{a} Eigenstates of the bulk (green), edge (blue), and corner (red) states in a finite size moir\'{e} superlattice ($5\times5$ primitive unit cell). \textbf{b} Maximal Wyckoff positions and the mirror reflection symmetry in the unit cell. The dashed line represents the primitive unit cell. The rotational center of the twist corresponds to the Wyckoff position 2b. LSWs in the innermost dodecagonal quasi-atoms form OAL2b. \textbf{c}. Energy eigenvalues in the finite size systems. Red square and yellow circle represent the corner and edge states respectively. Gray dots represent the bulk states. \textbf{d} Table representing the second Stiefel--Whitney number of each global band gap. The second Stiefel--Whitney number is determined by the number of the inversion-odd bands at the TRIM points. }
	\label{fig:3}
\end{figure*}

    In the followings, we show that LSW induced by the interference effect of the moir\'{e} pattern is characterized by non-trivial topology. Despite the complicated quasi-atom configurations, the moir\'{e} superlattice in this work possesses reflection symmetries $\mathcal{M}_x$ and $\mathcal{M}_y$ along the $\hat{x}-$ and $\hat{y}-$ axis (see Fig. \ref{fig:3}\textbf{b}). 
\bea
\mathcal{M}_x \: : \: \psi(x,y) \rightarrow \psi(x,-y)\ ,\ 
\:
\mathcal{M}_y \: : \: \psi(x,y) \rightarrow \psi(-x,y)\ .
\eea
    These symmetries are preserved regardless of the specific value of the commensurate twist-angles. The Bloch wave functions at the (high-symmetry) time-reversal invariant moment ($\mathbf{k}_*\in \{\Gamma,\textrm{M}^{0},\textrm{M}^{+},\textrm{M}^{-}\}$) are classified by the band representations according to the symmetry transformation properties under the little co-group at $\mathbf{k}_*$. Fig. \ref{fig:3} (b) shows the maximal Wyckoff position (WP) of the corresponding space group $Cmm2$ (No. 35). Interestingly, WP 2b corresponds to the rotational center of the moir\'{e} superlattice, where the Wannier function center of LSW is located. As a result, the flat band composed of LSW realizes the obstructed atomic limit with WP 2b (OAL2b). We explicitly confirm this feature, utilizing the method of the topological quantum chemistry \cite{Bradlyn2017, PhysRevResearch.1.032005}. By categorizing the irreducible representations and the corresponding band representations of all bands (see Supplementary Materials S5 for the explicit classification). We find that the bands separated by global gaps are characterized by the OAL2b if the bands contain the odd number of LSW.

    Furthermore, the OAL2b phase manifests as the non-trivial mirror-symmetry resolved Zak phase of the wave functions (see Supplementary Materials S5 for the rigorous proof). Since the system can be decomposed into even and odd $\mathcal{M}_x$ sectors along the reflection symmetric line $\Gamma-M-\Gamma$ in the BZ, we can separately define the bulk polarization of each sector as,
\bea
\nu_{\pm}= i\oint_{\mathbf{k}\in \Gamma-M-\Gamma} d\mathbf{k} \mathcal{A}(k_x,k_y)_\pm\ ,
\eea
    where cyclic integration is performed along the reflection symmetric line, $\Gamma-M-\Gamma$. $\mathcal{A}(k_x,k_y)_\pm=\langle \Psi_{\pm} (k_x,k_y)| \partial_k|\Psi_{\pm} (k_x,k_y)\rangle$  is the Berry connection defined for each reflection sector. $\Psi_{\pm} (k_x,k_y)$ is the Bloch wave function defined in the momentum space, which belong to the each symmetry sectors. The additional reflection symmetry, $\mathcal{M}_y$, further ensures $\mathbb{Z}_2$ classification of the bulk polarizations, $\nu_{\pm}\in \{0,\pi\}$~\cite{schnyder2008classification,PhysRevB.82.115120}. Evaluating the polarization for each band gap, we find that the band gaps separating the flat bands of the OAL possess non-trivial polarization, $\nu\equiv \nu_+ = \nu_-=\pi$, while the overall polarization, $\nu\equiv \nu_+ + \nu_-=0 \quad (\textrm{mod} \, 2\pi)$, has always trivial $\mathbb{Z}_2$ index. The non-trivial polarization with $\nu_{\pm}=\pi$ manifests as one-dimensional topological boundary modes. To explicitly show this feature, we consider a slab geometry with open boundary conditions along the $\hat{y}-$ direction; wave functions and their normal derivatives are continuous across the resonator boundary and fulfills Sommerfeld radiation conditions at the infinity. Figure \ref{fig:2}\textbf{d} shows that a pair edge spectrum emerges within the gapped region in the bulk bands. The two wave functions of Re$(\psi)$ shown in Fig.~\ref{fig:2}\textbf{e} correspond to symmetric and antisymmetric edge modes localized on each side of the boundary for $k_x=0$. 

    \paragraph*{Higher-order topology without reflection symmetry\\}

    The topological edge modes are robust against any arbitrary perturbations, preserving the underlying reflection symmetries. However, boundary terminations along an arbitrary direction naturally break the reflection symmetries. In such a case, the polarization, $\nu$, is ill-defined, and the topological robustness of the edge modes becomes deficient. Nevertheless, the product of the two reflection symmetries, which acts as an inversion, $\mathcal{P}=\mathcal{M}_x \times \mathcal{M}_y$, can be preserved. Such novel orientational dependence of the moir\'{e} pattern realizes a richer topological structure, characterized by the second Stiefel--Whitney number, $\nu_2$. We assign $\nu_2$ using the parity eigenvalues at time-reversal invariant momenta (TRIM) $\Gamma_i \in \{\Gamma,\textrm{M}^{0},\textrm{M}^{+},\textrm{M}^{-}\}$ as,
\bea
	(-1)^{\nu_2} =  \prod_{\Gamma_i\in{\rm TRIM} } (-1)^{[N^-(\Gamma_i)/2]},
	\label{eq:SW}
\eea
    where $N^-(\Gamma_i)$ is the number of bands with the odd parity ($\mathcal{P}=-1$) at $\Gamma_i$~\cite{Ahn_2019}. Figure \ref{fig:3}\textbf{d} shows the resulting symmetry classifications of each group of interconnected bands separated by global band gaps. The non-trivial second Stiefel--Whitney number manifests as topological corner modes in a finite size flake geometry with an open boundary (Red states in Fig. \ref{fig:3}\textbf{a}). We indeed find the mid-gap topological corner modes to be strongly localized at the inversion symmetric corners, in addition to being reminiscent of the one-dimensional edge modes (Blue states in Fig. \ref{fig:3}\textbf{a}). These topological corner modes remain robust against arbitrary perturbations as long as the inversion symmetry is intact. Another way to understand the origin of the corner mode is to consider the non-trivial polarization and the effective Su-Schrieffer-Heeger (SSH) model. The physical manifestation of the non-trivial polarization $\nu_\pm=\pi$ is the one-dimensional edge modes, as explained above only when the boundary respect the reflection symmetry. However, a generic edge termination is incompatible with the $\mathcal{M}_{x,y}$ symmetry, which gives rise to the energy gap in the corresponding edge spectrum. Finally, as one considers the finite size flake with a full open boundary condition (see, Fig.~\ref{fig:3}\textbf{a}), the corner between the two edges forms an effective SSH type domain wall, where a localized corner mode exists on the top and the bottom corners. (The mathematical proof of interconnection between the second Stiefel--Whitney number and the non-trivial polarization is given in Supplementary Materials.)

%\paragraph*{Discussion} 
\section*{DISCUSSION} 

    This work firstly demonstrates the strong interlayer coupling regime of a moir\'{e} superlattice of photonic crystals. Silicon (Si)-based materials have a typical refractive index of $3.5<n<7$ for a wavelength of $0.35~\mu m<\lambda<2~\mu m$, while that of semiconductor alloys such as GaAs, AlGaSb, and InGaAsP is $3.3<n<5$ for a wavelength of $0.35~\mu m<\lambda<2~\mu m$~\cite{rii,palik1998handbook}. Our demonstration can be experimentally realized in these materials.

    We also newly discover the emergence of topological flat bands at large twist-angles. This feature, which has not been observed in electronic systems, is the hallmark of our photonic moir\'{e} superlattice. The non-trivial topology of the moir\'{e} pattern realizes a tunable deformation of the one-dimensional edge modes and topological corner modes. Our topological material design using a moir\'{e} superlattice can be a promising start in the hunt for engineering a variety of topological photonic phases and for practical photonic device applications.

\section*{METHOD}

    The energy bands of the optical modes in the photonic crystals are obtained by solving the Maxwell equations reduced to the Helmholtz wave equation,
\bea
	-\nabla^2\psi=n^2(\mathbf{r})\frac{\omega^2}{c^2}\psi\ ,
	\label{eq:helm2}
\eea
    where $n(\mathbf{r})$ is the piecewise constant refractive index, and $\omega =c k$ is the free-space temporal frequency with vacuum wavenumber $k$ and speed of light $c$. An individual quasi-atom has a refractive index of $n=4$ and a radius of $r=a/6$ with a tunable $C_3$ rotation symmetric deformation, where $a$ is the original hexagonal lattice constant. Our results focus on the transverse magnetic [TM; $\psi = (0, 0, E_z )$] polarization of modes, where $E_z$ and $\vec{\nu}\cdot\nabla E_z$ are continuous at the interface between two different domains of the refractive index $n$. Here, $\vec{\nu}$ denotes an outward normal vector of the domain boundary. Note that \eq{eq:helm2} governs the transverse-electric [TE; $\psi = (0, 0, H_z )$] mode as well, but obeys a different boundary condition: $H_z$ and $1/n^2 \vec{\nu}\cdot\nabla H_z$ are continuous. Because of this different boundary condition, the Brewster angle (where reflection becomes zero) exists in the TE case, which results in the different cavity-cavity couplings from the ones in the TM case. Nevertheless, as the form of the Helmholtz wave equation, \eq{eq:helm2}, does not change, our symmetry analysis is also consistently valid in the TE case (see Supplementary Materials for details). To avoid an overflow of the domain mesh in resolving our complicated bilayer structure that would lead to poor computational performance~\cite{BEM_benchmark}, we employ the boundary element method~\cite{BEM_wiersig,BEM_knipp} to tackle the numerical computations of \eq{eq:helm2}. We calculate $\omega/c$ in the two-dimensional lattice, the finite size system, and the one-dimensional lattice by imposing, respectively, a two-dimensional periodic condition, a pure outgoing condition at infinity, and a mixed boundary condition. On top of the basic formalism of the boundary element method, we implement the block Sakurai--Sugiura method~\cite{BEM_bss_proof,BEM_isakari_bss_fmm,BEM_zheng_bss_fmm,BEM_sakurai_bss_gen,BEM_gao_bss_3d} to compute the photonic band structures. It is emphasized that all the computations in this work are conducted by our own implementations of the aforementioned numerical methods. Details of the numerical methods and the boundary conditions can be found in Supplementary Materials.

\clearpage
\newpage

%\bibliography{scibib}
\bibliographystyle{Science}

	\clearpage
    \newpage

\section*{Acknowledgments}
%Include acknowledgments of funding, any patents pending, where raw data for the paper are deposited, etc.
We acknowledge financial support from the Institute for Basic Science in the Republic of Korea through the project IBS-R024-D1.

\section*{Author information}
\paragraph*{Affiliations\\}
Center for Theoretical Physics of Complex Systems, Institute for Basic Science (IBS), Daejeon, 34126, Republic of Korea\\
Chang-Hwan Yi, Hee Chul Park, Moon Jip Park

%\paragraph*{Contributions\\}
%All authors of C.H. Yi, H. C. Park, and M. J., Park contributed this project by proposing this idea, initiating this project, collecting numerical data, analysing numerical data and carrying out mathematical analysis. All authors commented on and wrote the manuscript draft.

\section*{Corresponding author}
Correspondence to Hee Chul Park and Moon Jip Park.

\section*{Ethics declarations}
\paragraph*{Conflict of interest\\}
The authors declare no competing interests.

\section*{Supplementary materials}
Materials and Methods\\
Supplementary Text\\
Figures S1 to S16\\
Tables S1 to S4\\
References
\clearpage

\end{document}